# Electron-phonon interaction, excitations and ultrafast photoemission from doped monolayer $MoS_2$

Neha Nayyar, Duy Le, Volodymyr Turkowski, and Talat S. Rahman*

Department of Physics, University of Central Florida, Orlando, FL, 32816
*Corresponding author, e-mail address: talat@physics.ucf.edu

We analyze the effect of electron-phonon coupling on photoemission properties and ultrafast response of doped monolayer $MoS_2$. The analysis is based on combined DFT and many-body (Eliashberg theory) approaches. In particular, we have calculated the electronic and phonon spectra, the electron-phonon coupling and the electronic spectral function of the system at different values of doping. We have also analyzed the emissive properties and the response of the system to femtosecond (fs) laser pulses. It is shown that position of the emission peak of undoped system is in agreement with the experimental data if one takes into account the excitonic effects. The results for the self-energy and spectral functions of the doped systems suggest that one can expect ultrafast processes to be important in the system response , which makes the system attractive from the point of view of modern technological applications. Similar to graphene, the doped system demonstrates ultrafast (fs) relaxation of the electronic subsystem when excited by fs pulses, and a high ultrafast phonon relaxation-induced spectral fluence of visible light emission. Together with high carrier mobility, these features of monolayer $MoS_2$ might be used in modern optoelectronic technologies.

## I. Introduction

Single-layer $MoS_2$ shows a great potential for novel nanoelectronics,[1,2] nanooptics[3,4] and several other technologies. There several physical properties that make the system attractive from the technological point of view. In particular, they include a high electron mobility, room-temperature current on/off ratio and ultralow standby power dissipation. The last property makes the material attractive for field-effect transistors industry.[5] It was also shown that optical pumping of the system with circularly polarized light leads to complete (nanosecond-long) dynamic valley polarization in the system.[5,6] Other notable properties - tunable electronic structure and strong absorption and emission - make monolayer $MoS_2$ very attractive to be used in phototransistors,[7] light emitting devices operating in visible region[8] and other applications.

Strong luminescence, as opposed to its bulk counterpart,[9,10] is an especially striking phenomenon observed for this system. The intensity of the emission decreases about 100 times as the number of layers increases to bulk value. This decrease in intensity has been attributed to the indirect band gap in the bulk $MoS_2$. Many experiments have reported photoluminescence in single-layer $MoS_2$ and a great amount of work has been done in this field to understand the nature of the emission and absorption spectra.[10,11] Different experimental groups have reported slightly different values for the energy of the photoluminescence peak, ranging between 1.8 and 1.9 eV depending on the sample used.[9-14]

Absorption spectrum of the system also looks very attractive from the technological point of view, since the optical gap lays in visible range (~1.8eV) and the spectrum demonstrates peaks that correspond to strongly bound excitons[9] and trions.[15] Namely, it was found experimentally that the excitonic and trion binding energies are ~0.2-0.4eV[9,10,12] and 20meV,[15] correspondingly, which makes it possible to use these excited states in room-temperature applications. Several accompanying theoretical studies on the excitonis an trions monolayer $MoS_2$ have been performed. In particular, the phenomenological Wannier equation approach and the GW/Bethe-Salpeter equation (GW/BSE) approaches[16,17] were used to calculate the exciton energies and a trial wave function approach was used to study possibility of both the exciton and trion bound states in Ref. 18, while a TDDFT approach was proposed and used to calculate both exciton and trion binding energies in our work Ref. 19. All these studies confirm large binding energies for excitons and trions.

Another important property from the point of view of modern technologies is the ultrafast optical response of systems. It was recently found[20] that graphene, another monolayer material, demonstrates a strong emission when excited by ultrafast pulses. Following this pioneering work, several experimental studies on the ultrafast electron response[21-23] and on the ultrafast response together with time-resolved femtosecond emission[24,25] of grapheme were performed. Theoretical analysis[26-28] confirms the main experimental conclusion that the ultrafast electron response leads to an ultrafast emission, with dominant electron-phonon scattering of hot electrons.[26] Several other remarkable effects in excited graphene were predicted theoretically, [27,28] including the carrier multiplication and Auger recombination.

Naturally, one can expect a variety of attractive phenomena in excited monolayer $MoS_2$ as well, especially taking into account its strong absorbtive and emissive properties. Indeed, recently nonequilibrium response of monolayer $MoS_2$ structures has become a topic of active experimental studies.[29-33] Namely, ultrafast time-resolved photoemission experiments on single-layer $MoS_2$ flakes was performed in Ref. 29. It was shown that photocarrier recombination in these systems at low temperatures occurs on a few-picosecond timescale, while a biexponential photoluminescence decay was observed when the temperature increased. A femtosecond pump-probe pulse studies on a few-monolayer $MoS_2$ were performed in Ref. 30, where along with the values for the diffusion coefficient, diffusion length and mobility, it was estimated that carrier lifetime in this system is of order 100ps. Optical pump-probe spectroscopy studies[31,32] suggest that carrier cooling (dominated by phonon scattering), happens on picosecond scale, with 500fs electron thermalization, consequent 1-100ps,[31] defect-caused transient processes, and slower, ~10ns, recombination.[32] On the other hand, there are experimental indications of femtosecond response in the system: photoluminescence mapping femtosecond pump–probe spectroscopy measurements reveal a 50fs hole transfer from $MoS_2$ to $WS_2$ layer in photoexcited atomically thin $MoS_2/WS_2$ heterostructure.[33] Theoretical modeling predicts 100fs phonon lifetimes due to electron-phonon scattering. [34]

The results mentioned above are rather inspiring. Therefore, it is very important to understand theoretically in details the ultrafast response of monolayer $MoS_2$, including the photoluminescence properties. Theoretical analysis of the ultrafast response of optically excited systems is rather complicated problem. One of the main complications is necessity to take into account the electron-phonon scattering processes, which in many cases define the dispersion and the lifetime of excited electrons and, hence, the photoluminescence spectrum of the system. In this work, we perform such an analysis, together with studying of the absorption properties, and paying a special attention to the effects of electron-phonon interaction at different values of doping.

## II. Phonon properties

To calculate the phonon spectrum and the electron-phonon coupling parameters, as well as the ground state properties of isolated electronic system (with no phonon coupling) of monolayer $MoS_2$, we have used Quantum Espresso code.[35] The system consists of two planes of hexagonally arranged S atoms bonded through covalent bonds to central layer Mo atoms (Fig. 1). The lattice parameter was obtained by relaxing the system within the Local Density Approximation (LDA) XC potential (energy cut off 60 Ry, 15x15x1 k-point mesh). To avoid the effects of the interaction between images in the periodic unit cell calculations, the lattice parameter in the direction perpendicular to the monolayer plane was chosen to be 15 Å . The calculations gave the lattice parameter 3.167Å and the Mo-S bond length 2.419Å, in a reasonable agreement with experimental values 3.15Å and 2.414Å, respectively.[36,37]

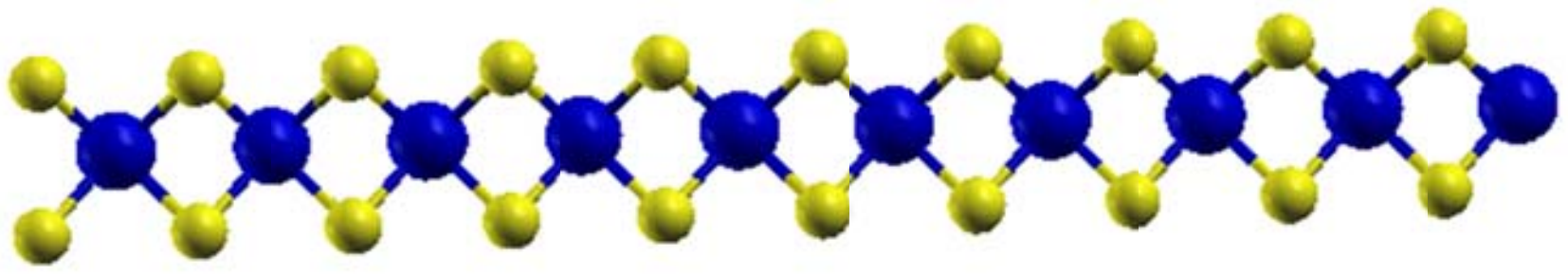

Figure 1. Schematic diagram of monolayer $MoS_2$ (blue – Mo atoms, yellow – S atoms).

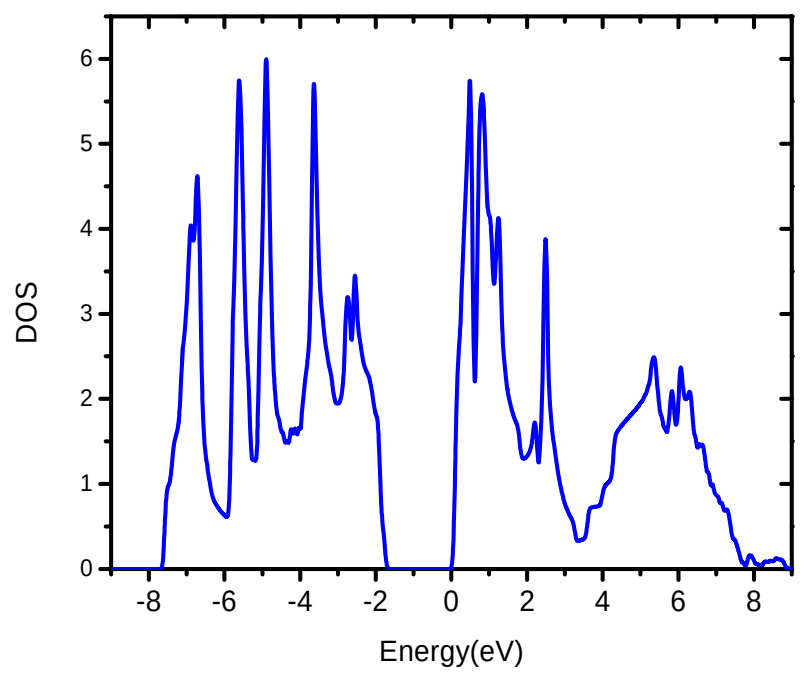

Figure 2. Electronic DOS of single-layer $MoS_2$.

The electronic density of states (DOS) of the system, shown in Figure 2, has the band gap of 1.8eV expected from other DFT calculations,[38] though it is smaller than experimental value 2.15eV.[39] It can be shown that the theoretical value of the gap can be increased by taking into account the GW correction ( though it seems that this approach leads to an overestimation of the gap, giving the value ~2.8eV[40]). On the other hand, the gap in the optical absorption spectrum (due to the excitonic effects) is equal 1.8eV. This result may be regarded as an accidental coincidence between the DFT electronic gap and experimental optical gap due to rather large excitonic binding energies, ~0.2-0.4eV (see, e.g., Ref. 19 and references therein). These large energies are caused in part by reduced screening of the electron-hole attraction in the two-dimensional systems. Another consequence of the reduced screening is rather long (up to 1000ps) exciton lifetime[41,42] and small (7-10 Å) exciton radius.[43] Below we discuss how the excitonic effects modify the optical spectra of the system in the undoped and doped cases.

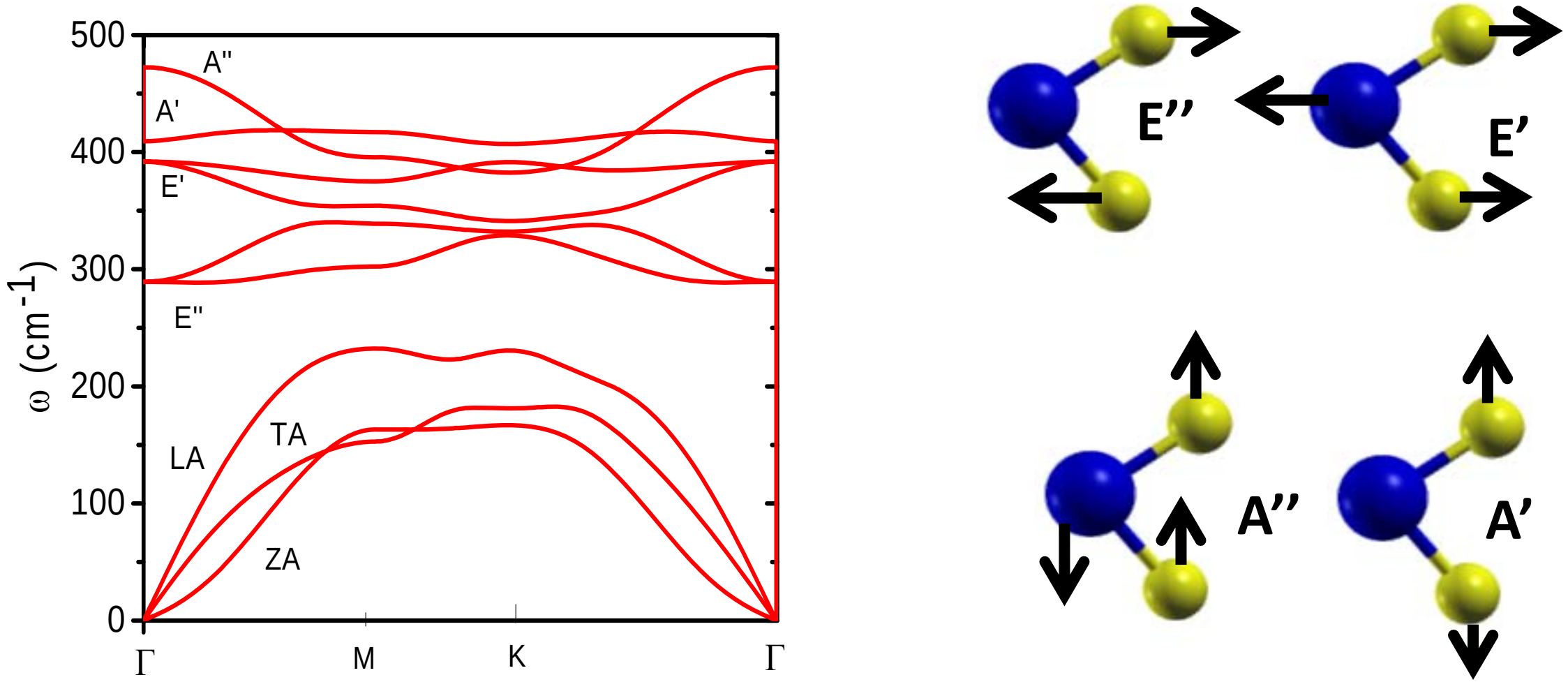

Figure 3. Phonon bandstructure (left) and schematic representation of the optical phonon modes (right) for monolayer $MoS_2$ obtained from DFT calculations.

Figure 3 shows the phonon dispersion and schematic representation of different optical modes of the spectrum. The system has the symmetry of point group D3h, which gives 9 branches of phonon curves, 3 acoustic and 6 optical modes. The acoustic modes LA and TA are in-plane vibration modes that have linear dispersion and are higher in energy than the out-of-plane vibration acoustic mode ZA. The lowest-energy optical modes $E'$ and $E''$ are in-plane vibrations, both degenerate at the Γ-point. Two $E''$ modes are vibrations with 2 sulphur atoms moving out-of-phase and Mo atom being static. The $E'$ modes are polar modes with Mo and S atoms moving out-of-phase with respect to each other. $A'$ is a homopolar mode with two sulphur atoms vibrating out-of-phase and fixed Mo atom, and $A''$ is the highest energy optical mode with out – of-plane vibrations, when Mo and S atoms are vibrating out-of-phase with respect to each other.
The results for the phonon spectrum presented in Fig.3 are in agreement with experimental data, namely the Raman spectra for the optical modes frequencies at Γ-point[50-54] (in particular, ~385$cm^{-1}$ for the $E_{2g}^1$ and 400-410$cm^{-1}$ for the $A_{1g}$ mode), as well as with other DFT calculations of the phonon spectrum in monolayer $MoS_2$.[55-59]

To calculate the electron-phonon coupling we have applied the Quantum Espresso post-processing software, which uses the results for the ground state atomic and electronic configurations, wave functions and bandstructure to obtain the electron scattering probability coefficients

$$g_{q\nu}(k,i,j) = \left(\frac{\hbar}{2M\omega_{q\nu}}\right)^{1/2} \left\langle \psi_{i,k} \left| \frac{dV_{SCF}}{d\hat{u}_{q\nu}} . \hat{e}_{q\nu} \right| \psi_{j,k+q} \right\rangle, \tag{1}$$

which correspond to the scattering of electron from state $i$ (momentum $k$) to state $j$ (momentum k+q) due to absorption (emission) of phonon with mode index $\nu$ and momentum q(-q)). In Eq. (1), $M$ is the atomic mass, $\psi_{i,k}$ and $\psi_{j,k+q}$ are the electronic wave functions for the initial and final states, respectively. $\frac{dV_{SCF}}{d\hat{u}_{q\nu}}$ is the gradient of the self-consistent potential with respect to the atomic displacements induced by the phonon mode $(\mathbf{q}, \nu)$ with frequency $\omega_{q\nu}$ and polarization vector $\hat{e}_{q\nu}$ (for a summary on the Eliashberg formalism to study the effects of the electron-phonon interaction used in this work, see, e.g., Ref. 44).

Next, from the functions (1) one can obtain the phonon line widths:

$$\gamma_{q\nu} = 2\pi\omega_{q\nu} \sum_{ij} \int \frac{d^3k}{\Omega_{BZ}} \left|g_{q\nu}(k,i,j)\right|^2 \delta(\varepsilon_{q,i} - \varepsilon_F)\delta(\varepsilon_{k+q,j} - \varepsilon_F), \tag{2}$$

where $\Omega_{BZ}$ is the volume of the first Brillouin zone, $\varepsilon_{q,i}$ is the energy of the electron in the state (band) $i$ and with momentum q, and $\varepsilon_F$ is the Fermi energy. $\gamma_{q\nu}$ defines the electron-phonon coupling constant for the corresponding phonon mode ν with the wave vector q:

$$\lambda_{q\nu} = \frac{\gamma_{q\nu}}{\pi\hbar N(\varepsilon_F)\omega_{q\nu}^2} \tag{3}$$

($N(\varepsilon_F)$ is the electron DOS at the Fermi level). Since the electron-phonon coupling parameters depend on the DOS at the Fermi level, in order to calculate them in the case of undoped semiconductors (with $N(\varepsilon_F) = 0$) one has to manually shift the Fermi level to a point where we have a finite DOS, i.e. introducing a doping into the system. In the most cases, the details of the electron-phonon interaction described by functions (2), (3) is not necessary, and one often uses the "averaged" Eliashberg function

$$\alpha^2F(\omega) = \frac{1}{2\pi N(\varepsilon_F)} \sum_{q\nu} \delta(\omega - \omega_{q\nu}) \frac{\gamma_{q\nu}}{\hbar\omega_{q\nu}}, \tag{4}$$

for this purpose. This function describes the scattering of an electron with fixed initial energy and momentum to all other states which differ in energy by $\hbar\omega$ (emission/absorption energy of the phonon that takes part in the scattering process). Furthermore, one can use even simpler function to describe the effect of phonons on the electronic subsystem - the effective (averaged) electron-phonon coupling parameter $\lambda$, which is connected with the Eliashberg function in the following way:

$$\lambda = \sum_{q,\nu} \lambda_{q\nu} = 2 \int \frac{\alpha^2F(\omega)}{\omega} d\omega. \tag{5}$$

The results for $\lambda$ in monolayer $MoS_2$ at different values of doping are presented in Table 1. We have obtained these results by shifting the Fermi level to a point with a finite DOS. In particular, the shift from zero of the conduction band by 0.067eV, corresponds to 1%-doping, or $8.64\times10^{14}$ $cm^{-2}$, the value of doping expected to be accessible in experiments. As it follows from Table 1, the coupling strongly depends on doping, though in the weakly-doped case one can estimate this parameter to be of order 0.1, in agreement with previous studies.[45]

Table.1 Electron-phonon coupling constant for different values of doping in monolayer $MoS_2$.

| Doping | Doping in $cm^{-2}$ | λ |
|---|---|---|
| 1% | $8.64\times10^{14}$ | 0.12 |
| 2% | $4.32\times10^{14}$ | 0.16 |
| 3% | $2.88\times10^{14}$ | 0.20 |

Below we shall use these values to study the optical properties of the system, assuming that the doping is not very large.

## III. The absorption and the emission spectra

To study the absorption and emission properties, one can calculate the momentum-resolved spectral function A(k,ω) (proportional to the probability for an electron to occupy the state with energy $\hbar\omega$ and momentum k):

$$A(k,\omega) = \frac{|Im\sum(k,\omega)|}{[\omega-\varepsilon_k-Re\sum(k,\omega)]^2+[Im\sum(k,\omega)]^2}, \quad (6)$$

where $\varepsilon_k$ is the band dispersion in the absence of interactions, $\sum(k,\omega)$ is the complex self-energy that describes the effects of many-body interactions, including coupling to phonons. One of the effects of the self-energy on the "free" electron spectrum is renormalization of the energy values (real part) and making the lifetimes of electrons finite (inverse of the imaginary part). The lifetime of the electronic excitation of the interacting system (frequency $\omega$) can be estimated as: $\tau(\omega) = -1/Im[\Sigma(\omega)]$. In the lowest-order quasi-elastic approximation, the phonon contribution to the real and imaginary parts of the electron self-energy is defined by the Eliashberg function as follows:

$$\mathrm{Re}\Sigma(E,T) = \int_{-\infty}^{\infty} d\nu \int_0^{\omega_{max}} d\omega' \alpha^2 F(E,\omega') \frac{2\omega'}{\nu^2-\omega'^2} f(\nu+E), \quad (7)$$

$$\mathrm{Im}\Sigma(\epsilon_i,k;T) = \pi \int_0^{\omega_{max}} \alpha^2 F^E(\epsilon_i,k;\omega)[1-f(\epsilon_i-\omega)+ f(\epsilon_i+\omega)+2n(\omega)]\, d\omega, \quad (8)$$

where $f$ and $n$ are the Fermi and Bose distribution functions, respectively, and $\omega_{max}$ is the maximal phonon (cut-off) frequency.

By using the results for the Eliashberg function and electron spectrum, we have calculated the electron self-energy and spectral function from Eqs. (6)-(8). The typical curve for the spectral function, at 1%-doping, is presented in Fig.4 (following the results for the phonon spectrum in Fig. 3, we have chosen the characteristic value $\omega_0 = 0.0497eV$ for the (optical) phonon frequency). In the insert of Fig.4 we show the corresponding frequency dependencies of the real and imaginary parts of the electron self-energy. As it follows from the inset, the values of the imaginary part of Σ at low frequencies (energies ~1eV) suggest that one needs to take into account femtosecond processes (scattering times $\sim 1/\mathrm{Im}\Sigma$) in order to properly describe the ultrafast dynamics in doped monolayer $MoS_2$. It must be noted that the carrier lifetimes 1-100ps, reported in literature[29-32] in the case of undoped excited system, most probably are related to the exciton recombination times, which can be caused by different reasons. The doped system may demonstrate faster dynamics, due to enhanced electron-phonon interaction. We show some of the consequences of this interaction in the following Section. Other pronounced features of the solution for the electron self-energy in Fig.4 is suppressed imaginary part of Σ at frequencies smaller than the phonon frequency, $|\omega| < \omega_0$, due to the carrier occupancy-related constraints on the scattering processes, and the peaks of the real part of the self-energy at $\omega \sim \pm\,\omega_0$ due to enhanced electron-phonon scattering processes. Interestingly, similar frequency-dependence for the self-energy in graphene with a linear free-carrier spectrum results in a kink in the electronic spectra function at frequencies close to $-\omega_0$ (see, e.g., Ref. 61).

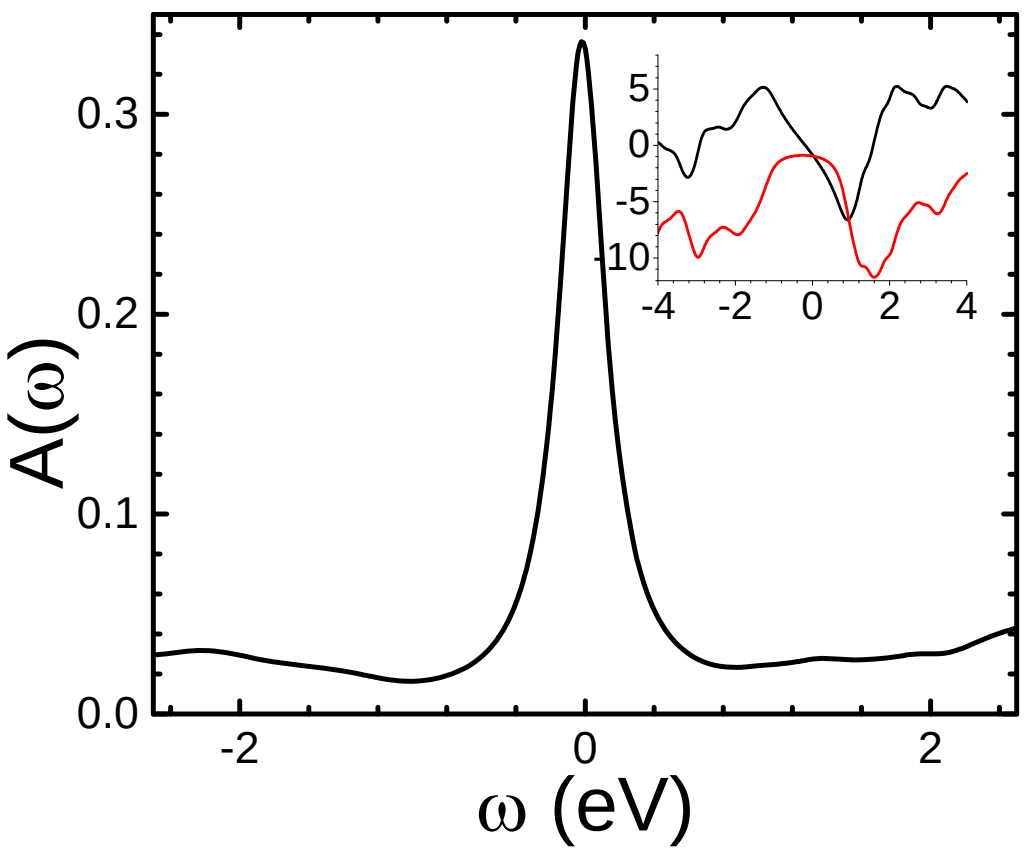

Figure 4. The spectral function of monolayer $MoS_2$ at doping 1%. Inset: the corresponding real (black) and imaginary (red) parts of the electron self-energy.

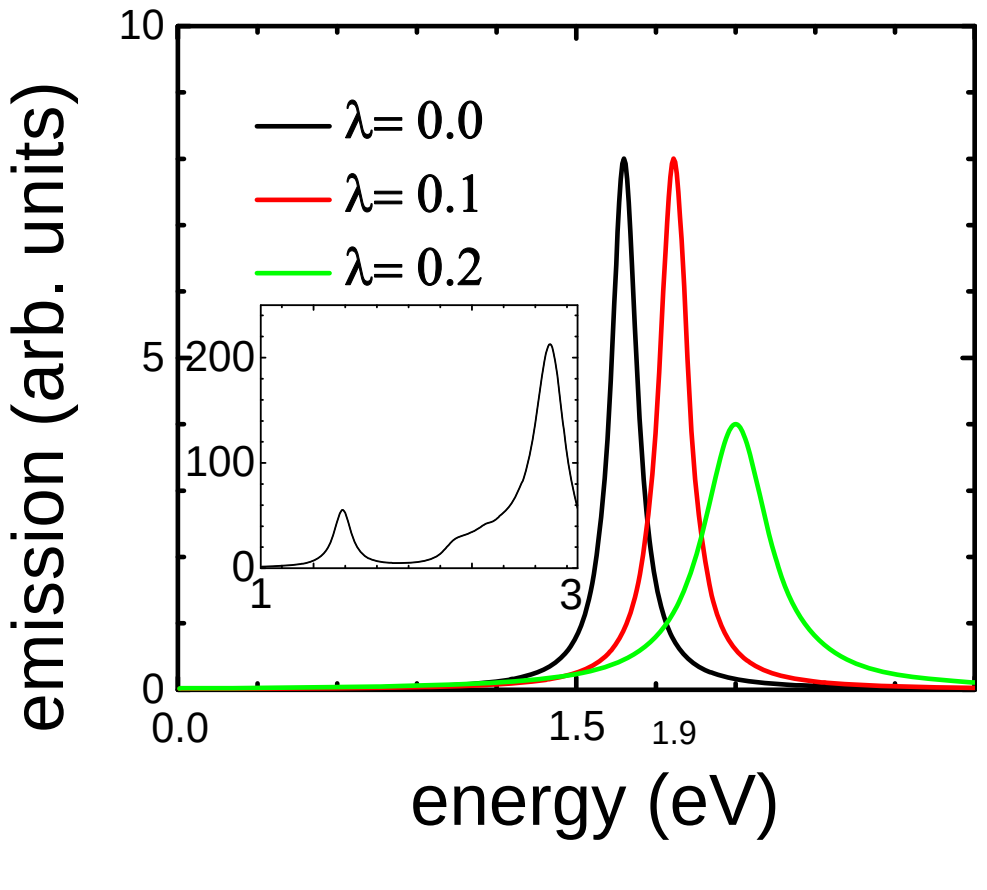

Figure 5. The emission spectrum of monolayer $MoS_2$ at different values of $\lambda$. Insert the corresponding absorption spectrum at λ=0.

The results for the absorption and emission spectra (at different $\lambda$'s) of the undoped system are presented in Fig.5. The absorption spectrum and emission spectra were calculated by including excitonic states within density-matrix TDDFT[46] with the Slater exchange-correlation kernel (see also Ref. 19). We intentionally have shifted the spectra to the right by 1eV, by correcting our DFT bandgap towards the experimental values (according to more accurate theoretical calculations, see above). The absorption spectrum shows two prominent features that correspond to the exciton transition (binding energy ~1.0eV) and the valence-conduction band excitations (only lowest conduction band has been taken into account). The results shown in Fig.5 suggest that the system is optically active in the visible range (though the spectra should be slightly shifted to the right ).

Here it must be mentioned another type of the excitations which might take place in doped system: excited electron-hole bound states of Mahan excitons (MEs).[47] The possibility of these excitations in doped systems is actively discussed in the literature nowadays (see, e.g., Ref. 48). The ME excitations are expected to brcome suppressed with doping increasing due to screening of the electron-hole attraction by conduction band charges. The Mott criterion for the existence of the ME states $q_{TF}a_X > 1.19$, where $q_{TF}$ is the Tomas-Fermi wave vector and $a_X$ is the exciton radius, makes them suppressed in the 3D case at rather low values of doping. Since in the 2D case the Thomas-Fermi vector is doping-independent (in the case of parabolic dispersion at not very large doping) and equal to $2/a_B$ ($a_B$ is the Bohr radius), and the excitonic radius in monolayer $MoS_2$ is estimated as $a_X$~7-10A, one can expect the Mott criteria to be satisfied for rather high values of doping (when the simplified formula for the parabolic dispersion remains correct). The question of the lifetime of the ME in this system remains open so far.

The emission spectrum is obtained from the spectral function by taking into account the phonon correction to the electron self-energy. The PL peak position depends on the electron-phonon coupling, the peak moves to higher energies with the value of λ increasing and as expected the linewidth also increases with the value of λ increasing. The position of the PL peaks for all three values of λ are in a reasonable agreement with experimental values[29,30] (again, with taken into account the renormalization of the bandgap).

# IV. Ultrafast response of doped system

Finally, we apply a rather simple analysis in order to consider a possibility of the ultrafast response of the system to an external laser perturbation. As it was already mentioned in the Introduction, ultrafast experimental studies of undoped $MoS_2$ systems (flakes,[29] few-monolayer,[30] multilayer,[31] monolayer,[32]) reveal picosecond response, with the general conclusion that most probably this timescale is defined by the exciton recombination time. Indication of faster processes were found in experimental work 33, where the authors reported 50fs hole transfer from a $MoS_2$ to $WS_2$ layer of the corresponding heterostructure, and in theoretical paper by Kaasbjerg et al.,[34] who predicted phonon lifetime of 100fs in monolayer $MoS_2$. On the other hand, the case of doped system was not analyzed so far, and it would be important to explore a possibility of the ultrafast response in metallic $MoS_2$. Femtosecond response in doped graphene has been reported for the first time in paper 20, where it was demonstrated that the electron relaxation happens in tens of femtosecond regime, with a similar scale for the consequent irradiance times. Since the electron-phonon coupling parameter and several other characteristics of grapheme and monolayer $MoS_2$ are similar, and the emissivity (in the undoped case) is even much stronger for $MoS_2$, one might expect pronounceable emission and other ultrafast processes in the last system as well.

To explore such a possibility, we begin with establishing possible ultrafast time-scales in the doped system. For this, we consider the equilibration (thermalization) of the electronic subsystem due to the electron- phonon scattering by using the Allen approximation,[49] in which the equation for the time dependence of the electronic temperature $T_e$ has the following form:

$$\frac{dT_e}{dt} = I(t) + \gamma_T(T_L - T_e), \tag{9}$$

where $T_L$ is the lattice (room) temperature, $I(t)$ is the time-dependent external pulse field and

$$\gamma_T = \frac{3\hbar\lambda\langle{\omega_{ph}}^2\rangle}{\pi k_B T_e} \tag{10}$$

is the relaxation rate ($\langle{\omega_{ph}}^2\rangle$ is the averaged square of the phonon frequency). Since there are two parameters in the model, electron-phonon coupling $\lambda$ and phonon frequency $\sqrt{\langle{\omega_{ph}}^2\rangle}$, we performed calculations using the values for these parameters in the range obtained from DFT in the low-doped case: the coupling constant between 0.1 and 0.2 and phonon frequency between 0.01eV and 0.75eV. As it follows from our calculations, one can expect a femtosecond dynamics for this range of parameters (Fig.6).

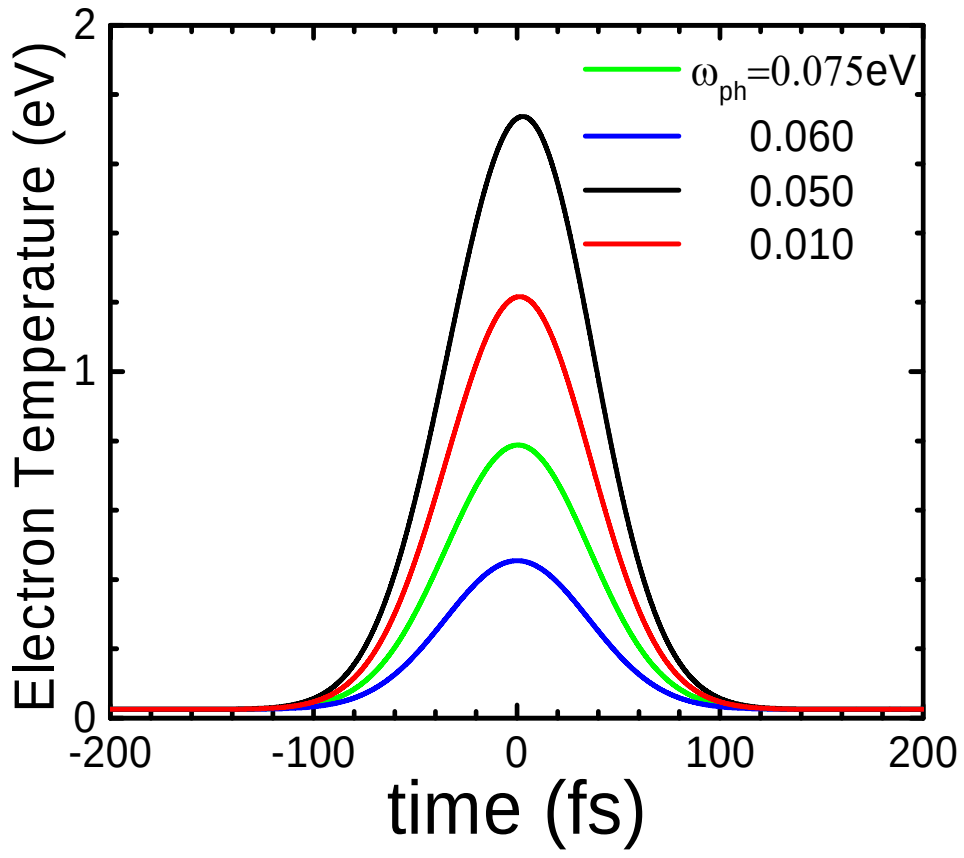

Figure 6. Time dependence of electron temperature in doped $MoS_2$ due to phonon scattering at different values of the phonon frequency and λ=0.12. The applied external pulse has Gaussian shape with the center at t=0s and the width 100fs.

Since the time of relaxation of the electronic subsystem to the room temperature $T_L$ defines the timescale of the equilibration of the whole system, i.e. the times when also all the emission processes happen, one can expect the phonon-stimulated radiation also happen in this regime. We analyzed the energy dependence of the spectral fluence of the system (the total radiant energy emitted in all directions per unit area per photon energy) by using simple Planck formula:

$$F(\omega, T) \sim \frac{\tau_{em}\varepsilon(\omega)\omega^3}{\left[e^{\frac{\omega}{T}}-1\right]}, \tag{11}$$

where $\tau_{em}$~100 fs is the emission time (the value chosen by using the results of Fig. 6), $\varepsilon(\omega)$ is the frequency-dependent emissivity, which needs to be determined experimentally, and for which we use the estimated for the visible range value $\varepsilon$~0.1 (obviously, the absolute value for this quantity only leads to a renormalization of the magnitude of the spectral fluence). In Fig.7 we present the energy dependence of the fluence at different values of the temperature of the system. As it follows from Figs. 6 and 7, one can expect ultrafast emission from the doped monolayer $MoS_2$, similar to graphene. Indeed, the temperature between the values shown in Fig.7 changes on the fs scale (Fig. 6), and the corresponding fluence at these temperatures (and for these times), must also have similar time dependence.

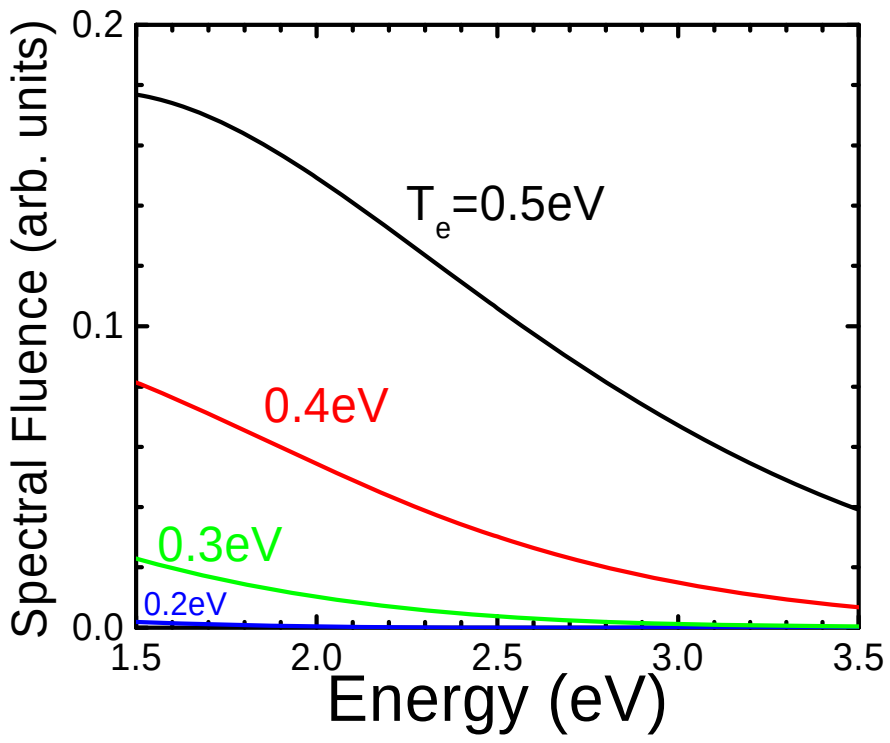

Figure 7. The spectral fluence of doped monolayer $MoS_2$ at different values of temperature. The values for other parameters are given in the text.

These results along with experimentally observed high electron mobility suggest rich ultrafast dynamics of the doped $MoS_2$, with possible practical ultrafast applications. Several interesting phenomena may exist, like carrier multiplication[27] and Auger recombination[28] predicted for graphene, and some other effects due to unique characteristics of $MoS_2$, first of all pronounced emissivity.

## V. Conclusions

To summarize, we have studied the effects of the electron-phonon coupling on the optical response of pure and doped monolayer $MoS_2$, including the ultrafast response. The absorption spectrum of undoped system demonstrates an excitonic peak in the visible range, in agreement with experimental data. The position of the emission peak is sensitive to the value of electron-phonon coupling, but it lays in the visible range, also in the experimentally observed range 1.8-1.9eV . Our calculated values of electron-phonon coupling constant are of order or even larger than the one obtained theoretically for doped graphene.[60,61] This suggests that similar to the last system,[20] one may expect ultrafast response of monolayer $MoS_2$ caused by thermalization effects. Indeed, the results for the electron self-energy of the doped systems gives estimated lifetimes of the electronic excitations on the femtosecond scale, and suggest that one can expect 10-100fs processes during the response to ultrafast excitations. Our study of the nonequilibrium ultrafast confirms this, showing the fs relaxation of electronic subsystem, and following visible phonon-assisted emission (fluence).

## Acknowledgements

We would like to thank the US Department of Energy for a financial support through Grant number DE-FG02-07ER46354.